# First-principles calculations of the dispersion of surface phonons of the unreconstructed and reconstructed Pt(110)

Sampyo Hong and Talat S. Rahman*

Department of Physics, Cardwell Hall, Kansas State University, Manhattan, KS 66506, USA

Rolf Heid and Klaus Peter Bohnen

Forschungszentrum Karlsruhe, Institut fuer Festkoerperphysik, D-76021 Karlsruhe, Germany

Abstract

We present result of calculations of the surface phonon dispersion curves for Pt(110) using density functional theory in the local density approximation and norm conserving pseudopotentials in a mixed-basis approach. Linear response theory is invoked and both the unreconstructed, and the missing row (1x2) reconstructed surfaces are considered. We find that the reconstruction is not driven by a phonon instability. Most of the observed phonon modes for the (1x2) structure can be understood in terms of simple folding of the (1x1) Brillouin zone onto that for the (1x2) surface. Largest changes in the phonon frequencies on surface reconstruction occur close to the zone boundary in the (001) direction. Detailed comparison of atomic force constants for the (1x1) and the (1x2) surfaces and their bulk counterparts show that the bulk value is attained after three layers. Our calculations reproduce nicely the Kohn anomaly observed along the (110) direction in the bulk. We do not find a corresponding effect on the surface.

## I. INTRODUCTION

In the area of surface science two experimental techniques, electron energy loss spectroscopy[1,2] and He atom-surface scattering[3], have produced much of the anticipated data on the dispersion of surface phonons of several metals. From the frequencies of the modes at the high symmetry points and their dispersion along specific directions in the surface Brillouin zone conclusions have been drawn about the changes in the nature of the bonding between the surface atoms as compared to those in the bulk. However, the complexity of the many-body interactions that control atomic vibrations require a detailed parameter free analysis which could be obtained only for a limited number of cases. First principles electronic structure calculations, for example, were feasible mostly for the frequencies of modes at the high symmetry points.[4] The majority of the theoretical discussion of the surface phonons centered around the usage of various types of force constants to fit the experimental data, leading to a good bit of debate about their validity and implications. The situation has changed with the availability of more efficient computational codes and enhanced computer power, as documented by a recent review article[5] in which the dispersion of the phonons of the low Miller index surfaces of several transition and noble metals have been presented. The work presented here is an off-shoot of what appears in the review article.  A quick look at Ref. 5 will show that while there are some genuine features in all dispersion curves, the details are specific to the elemental metal and of course, the surface geometry. Our interest in this paper is to present a detailed analysis of the changes in the force fields for the unreconstructed (1x1) structure of Pt(110) as well as for the missing row (1x2) phase. The focus on Pt surfaces stems from their relevance to catalysis and material science, in general. A comparison of the structural and dynamical properties of the reconstructed and unreconstructed surface may provide some insights into why the surface reconstructs. At the very least it would settle the issue of whether such informations may be obtained from phonon dispersion curves. Questions of a phonon anomaly as an extension of one in the bulk have also been raised for Pt(111).[6 – 9] Our effort here is to check also whether such an anomaly is present on the Pt(110) surface based on first principles electronic structure calculations. Thus while the dispersion curves for unreconstructed Pt(110) already appeared in Ref 5, this work contains a detailed analysis of the force constant field of the bulk and surface (reconstructed/unreconstructed) for this paper. The dispersion curves for the reconstructed surface are original to this work. In the next section we provide a summary of the calculational technique. This is followed in section III of the results and discussions. Conclusions are presented in section IV.

## II. DETAILS OF THEORETICAL CALCULATIONS

First principles calculations of the total energy of the system were performed using the mixed-basis pseudopotential method.[10] To represent ion-electron interaction a norm-conserving pseudopotential for Pt was used in the local density approximation (LDA) for electron-electron interaction,[11] a Hedin-Lundqvist form of the exchange-correlation functional was employed.[12] For the valence states of Pt, d-type local functions at each Pt site, smoothly cut off at a radius of 2.1 a.u., and plane waves with a kinetic energy of 16.5 Ryd were applied. Integration over the irreducible Brillouin-zone were carried out using a number of special kpoints (12 ~ 24 k-points). A Fermi level smearing of 0.014 Ryd was employed. Figure 1(a) shows two types of surface structures: (1x1) unreconstructed and (1x2) missing-row reconstructed (110) surfaces. For simulating the (110) surfaces, supercells containing eleven layers (11 Pt atoms for (1x1) and 20 atoms for (1x2)) were used with inversion symmetry. Structure relaxation was carried out by calculating forces using Hellmann-Feynman theorem until forces on all atoms were less than 0.001 Ryd/a.u. The calculations of phonon dispersions were performed using linear response theory within a mixed-basis perturbation theory approach.[13] Phonon dispersion curves were obtained by standard Fourier interpolation method using a (4x6) q-point mesh for the (1x1) and a (2x6) q-point mesh for the (1x2). Surface force constants were combined with bulk force constant by adding an asymmetric bulk slab of 100 layers to obtain projected bulk phonon modes. (As a result, one side of the slab is an ideal bulk-terminated surface.) For the bulk phonon calculation an (8x8x8) q-point mesh was used. Surface modes were identified by a weight larger than 20% in the first two layers.

## III. RESULTS AND DISCUSSIONS

### A. Bulk phonon dispersion

Our calculated bulk phonon dispersion curve is shown in Fig. 2. In Fig. 2(a) whole dispersion curve along the significant symmetry points in BZ along with the available experimental result[14] while the portion of the dispersion curve for (110) direction is separately shown in Fig. 2(b) to emphasize the existence of the weak bulk anomaly. The two bottom branches are TA modes and the upper one is a LA mode. Overall the calculated phonon frequencies are in excellent agreement with experiment, even describing all anomalous features along (110) direction. The very good sampling of the Fermi surface integration which was needed for obtaining the anomalous features required up to 8000 k-points in the whole BZ. The anomaly is due to Fermi surface nesting which is associated with transitions along (110) direction across the parallel portion of the Fermi surface formed by the fifth bulk band.[15] According to the previous force constant parameterization method,[8] six nearest neighbor (NN) couplings were necessary to correctly produce the weak anomaly. Our analysis of the first principles force constants shows that the anomaly is due to couplings well beyond second NN neighbors in agreement with the above mentioned parameterization. The important force constants relevant for the anomaly are very small (< 10% of NN coupling) and lead thus only to a weak anomaly.

### B. Surface phonon dispersions

Before discussing in detail the surface phonon dispersion for Pt(110) we would like to emphasize that all our calculations are performed for fully relaxed structures. For completeness we provide in Table I the relaxation parameters for the (1x1) and the (1x2) structures of Pt(110). The results are in agreement with other calculations as well as experimental data.[16] For later discussions, we present in Figs. 1 and 3 the top, as well as, the side view of the (1x2) structure with labeling of the different inequivalent atoms. One should keep in mind that the (1x1) surface is described by 1-atom per layer, while for the (1x2) surface there are two inequivalent atoms per layer which for symmetry reasons for odd numbered layers lead to the experimentally observed buckling. In Figure 4 we have plotted the phonon dispersion for the unreconstructed surface. Note that this dispersion curve can be found in Ref. 5. However, for comparison with those for the (1x2) structure, it is important that we include them here and present them such that the results are mapped onto the reduced BZ compatible with the (1x2) structure, with the dashes and the white circles representing the mapped bulk and surface states, respectively. We find in Fig. 4 a very rich spectrum of surface states and surface resonances especially along Γ-X and X-S' directions which is directly a consequence of the mapping onto a reduced BZ. The dispersion along Y'-Γ can be understood simply in terms of the doubling of that along Y- Γ as a consequence of the BZ folding in the (001) direction for the (1x2) structure.

Figure 5 shows the results for the (1x2) missing row structure. Along Γ-X and Y'-Γ the number of surface modes and resonances is remarkably reduced due to the fact that symmetry breaking at the surface allows the coupling of some pure surface modes for the (1x1) structure to bulk modes thus leading to a decay. Most of the remaining modes are still very similar in character to those for the (1x1) structure. At the Γ point all surface modes have disappeared compared with the results of the unreconstructed surface. At the X-point the biggest change occurs for the Rayleigh wave frequency and a high frequency mode around 20 meV. Other modes either hardly change their frequency unless they are one of the ones that decay, as discussed above.  Along the S'-Y'-direction the difference between the (1x2) and the (1x1) structure is most apparent. As a result of the symmetry breaking in this direction, formerly degenerate modes split and acquire very different frequencies leading to a rich scenario of true surface modes between 17 and 23 meV at the S'-point. The low lying modes have also been substantially modified and stiffened. At the Y' point, the very pronounced surface resonance around 20 meV has to be mentioned which is mostly polarized along <110> and which is absent on the unreconstructed surface. At the lower end of the spectrum the Rayleigh mode frequency is hardly modified however some new modes appear close to Y'. It is worth noticing that there is no instability seen either in the (1x1) nor in the (1x2) spectrum, thus no indication of strong electron phonon coupling leading to the reconstructed phase could be found. Our calculations also do not indicate any apparent anomaly at the surface to be related to the observed bulk phonon anomaly along (110). This is not surprising since this anomaly required a very special nesting feature of the Fermi surface which might not be present at the surface; furthermore matrix element effects might also influence this anomaly strongly since already in the bulk it was not very pronounced. Having discussed the behavior of the phonon dispersion curves we will now concentrate on a detailed comparison of the atomic force constants that emerge form our first principles calculations for the (1x1) and (1x2) surfaces, and those in the ideal bulk structure.

### C. Analysis of bulk and surface force constants

In Table II we present the force constants between atoms in the bulk up to the third nearest neighbors. While these numbers are informative, we need to find a more compact notation with which we can examine signatures of very long range interactions amongst the atoms. For such a purpose, it is useful to employ a condensed notation for measuring the coupling strength. We do this through an average force constant $I_{ij}$ between atoms i and j, defined as

$$< I_{ij} > \equiv \sqrt{\frac{1}{3}\sum_{\alpha\beta}(\Phi^{ij}_{\alpha\beta})^2}$$

where $\Phi^{ij}_{\alpha\beta}$ are the individual force constant component along the Cartesian coordinates α and β. The results for the average bulk force constant, up to the ninth nearest neighbor, are summarized in Table III. We note that there are basically three groups of parameters: the force constants between the first NN which are obviously the most dominant, those between the $2^{nd}$ to the $4^{th}$ NN which are roughly one order of magnitude smaller than the former, and finally those between the $5^{th}$ to the $9^{th}$ nearest neighbors which are another order of magnitude smaller. As we have already commented, the presence of interactions up to the $9^{th}$ nearest neighbor makes the case of bulk Pt distinct from that of metals like Ni and Cu for which interactions decay faster.[5] This slow decay for the case of Pt results from Fermi surface nesting and ultimately accounts for the weak Kohn anomaly in the bulk phonon spectrum. We now turn to Tables IV and V for a summary of the calculated force constant matrix elements between nearest neighbors atoms in our supercell. The intralayer force constants in Table IV naturally reflect a single bond length of 5.26229 a.u., while surface relaxations, buckling, and pairing presented in Table I account for the range of bond lengths that appear in the interlayer force constant matrices in Table V. As compared to the force constants between NN atoms in the bulk, as shown in Table II, we find the changes to be dominant in the first two layers on the unreconstructed Pt(110) and in the first three layers, on the reconstructed Pt(110) because of the missing-rows in the first layer. When we compare the force constant matrices of the unreconstructed Pt(110) with those of the reconstructed surface, a characteristic signature for reconstruction is not found. Interlayer force constant matrices are shown in Table V. First of all, the changes of the first NN average force constants of the unreconstructed and the reconstructed Pt(110) are perfectly consistent with the relaxations in Table I. Buckling of the third and fifth layer in reconstructed Pt(110) is also well described by two sets of the average force constants ((2,3)-(2,3') and (4,5)-(4,5')). Other than that, force constants of the unreconstructed and the reconstructed Pt(110) are very similar. Figures 6 and 7 show, respectively, the change of intralayer and interlayer average force constants with bond length for the unreconstructed Pt(110), while Figure 8 and 9 present the same for the reconstructed Pt(110). As in bulk, the force constants in Figs. 6 - 9 decrease by one order of magnitude for bond length larger than 7.442 a.u. and at least two orders of magnitude smaller for bond length larger than 12.8899 a.u.. Here 12.8899 a.u. may be regarded as an effective range of layer-layer interaction and gives us a measure of the minimum thickness of slab or vacuum needed in a supercell calculation to prevent surface-surface interaction. On the other hand, as compared with those for bulk, the intralayer average force constants for the first NN bond length in Figs. 6 and 8 are softened while the interlayer average force constants for the first NN bond length in Figs. 7 and 9 are stiffened. The intralayer average force constants for the reconstructed surface shows more softening than those for the unreconstructed surface while the interlayer average force constants for the reconstructed surface are similar to those for the unreconstructed surface.

## IV. CONCLUSIONS

We have presented here calculated surface phonon dispersion curves for Pt(110) using density functional theory in the local density approximation and the norm conserving pseudopotential method in a mixed-basis approach. Linear response theory is invoked and both the unreconstructed phase, as well as, the missing row (1x2) reconstructed surface are considered. While several interesting features are found in the phonon dispersion curves, we do not find the presence or indication of dramatic phonon anomalies which could have pointed to rational for phonon driven surface reconstruction. Most of the observed phonon modes for the (1x2) structure can be understood in terms of simple folding of the (1x1) Brillouin zone onto that for the (1x2) surface. A detailed comparison of force constants between the (1x1) and the (1x2) and their bulk counterparts shows that the bulk value is attained after three layers. While our calculations reproduce nicely the Kohn anomaly observed along the (110) direction in the bulk, we do not find a corresponding effect on the surface.

**ACKNOWLEDGEMENTS**

This work was supported in part by the National Science Foundation, under grant number CHE 0205064. Calculations were carried out both at NCSA, Champaign-Urbana, and at the facility at Forchungszentrum, Karlsruhe. TSR also thanks the Alexander von Humboldt Foundation for the award of a Forschungspreise which has facilitate the collaboration together with support from the International Division of NSF.

---

*Corresponding author. E-mail address: rahman@phys.ksu.edu; WWW homepage: http://www.phys.ksu.edu/~rahman/; FAX: 785 532 6806.

Figures

Fig. 1. (a) Surface structures of Pt(110) and (b) the corresponding surface Brillouin zone.

Fig. 2. Calculated Pt bulk phonon dispersion curves: (a) Along the high symmetry points. (b) Along the (110) direction. The circles are the neutron scattering data in Ref. 14.

Fig. 3. Side view of the (1x2) structure of reconstructed Pt(110).

Fig. 4. The phonon dispersion for the (1x1) structure.

Fig. 5. The phonon dispersion for the (1x2) missing row structure.

Fig. 6. Intralayer average force constants for unreconstructed Pt(110).

Fig. 7. Interlayer average force constants for $1^{st}$ to all.

Fig. 8. Intralayer average force constants for reconstructed Pt(110).

Fig. 9. Interlayer average force constants for $1^{st}$ to all.

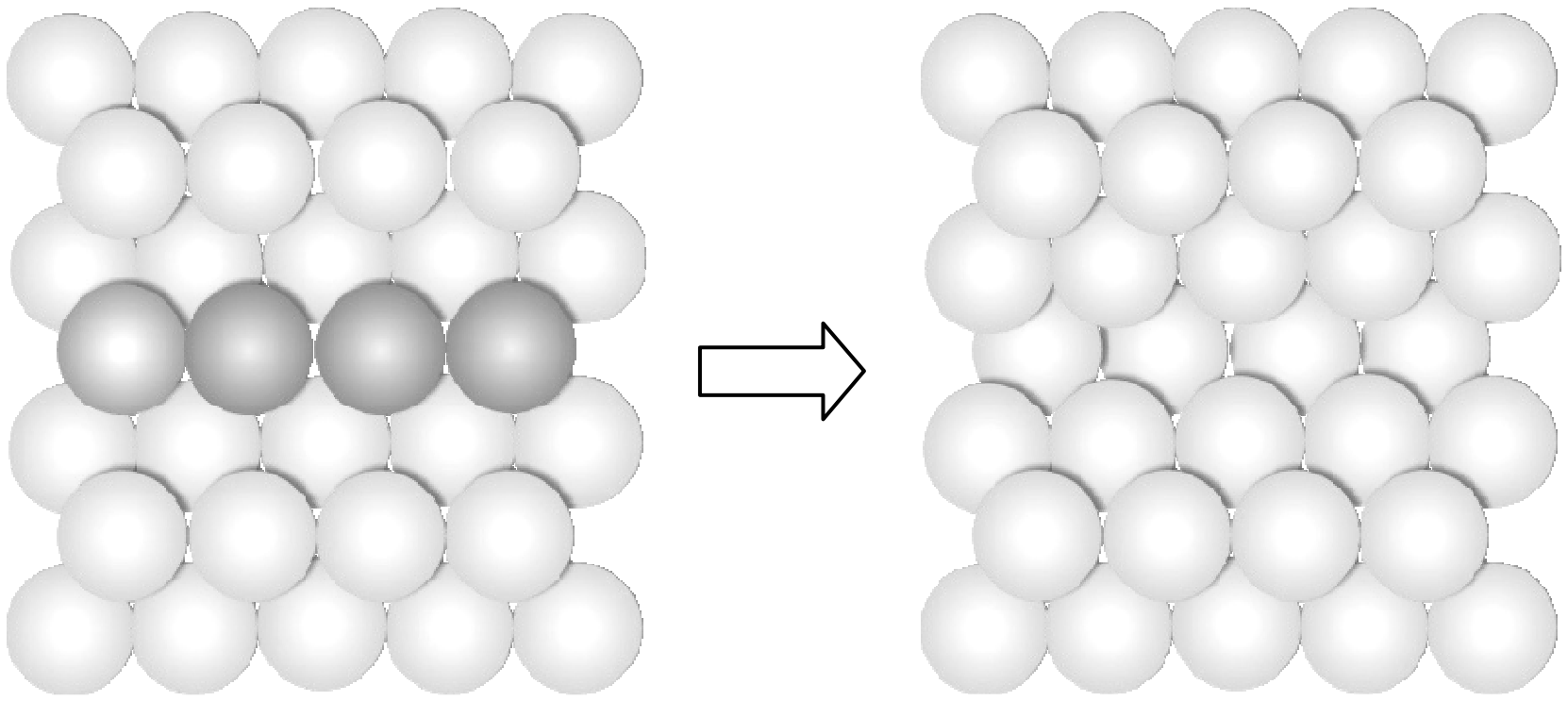

**Unreconstructed Pt(110)**

**(1x2) missing-row Reconstructed Pt(110)**

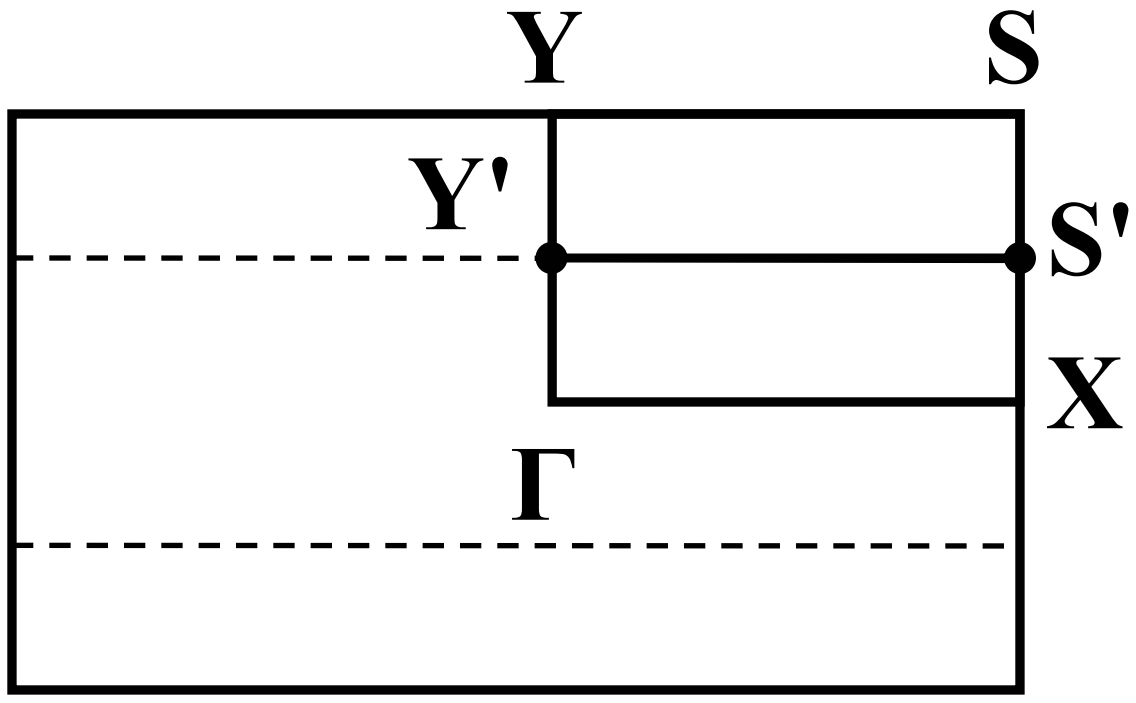

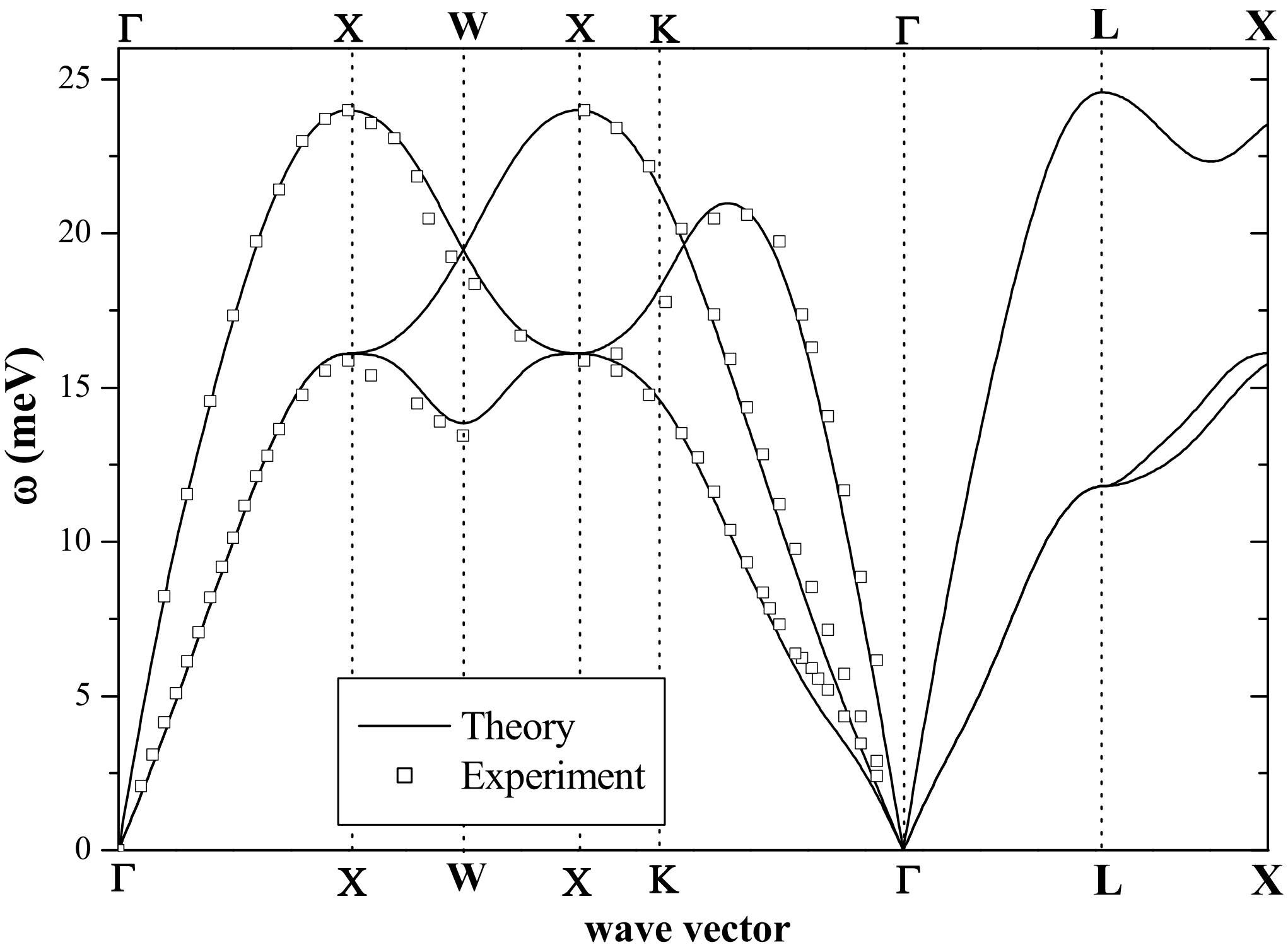

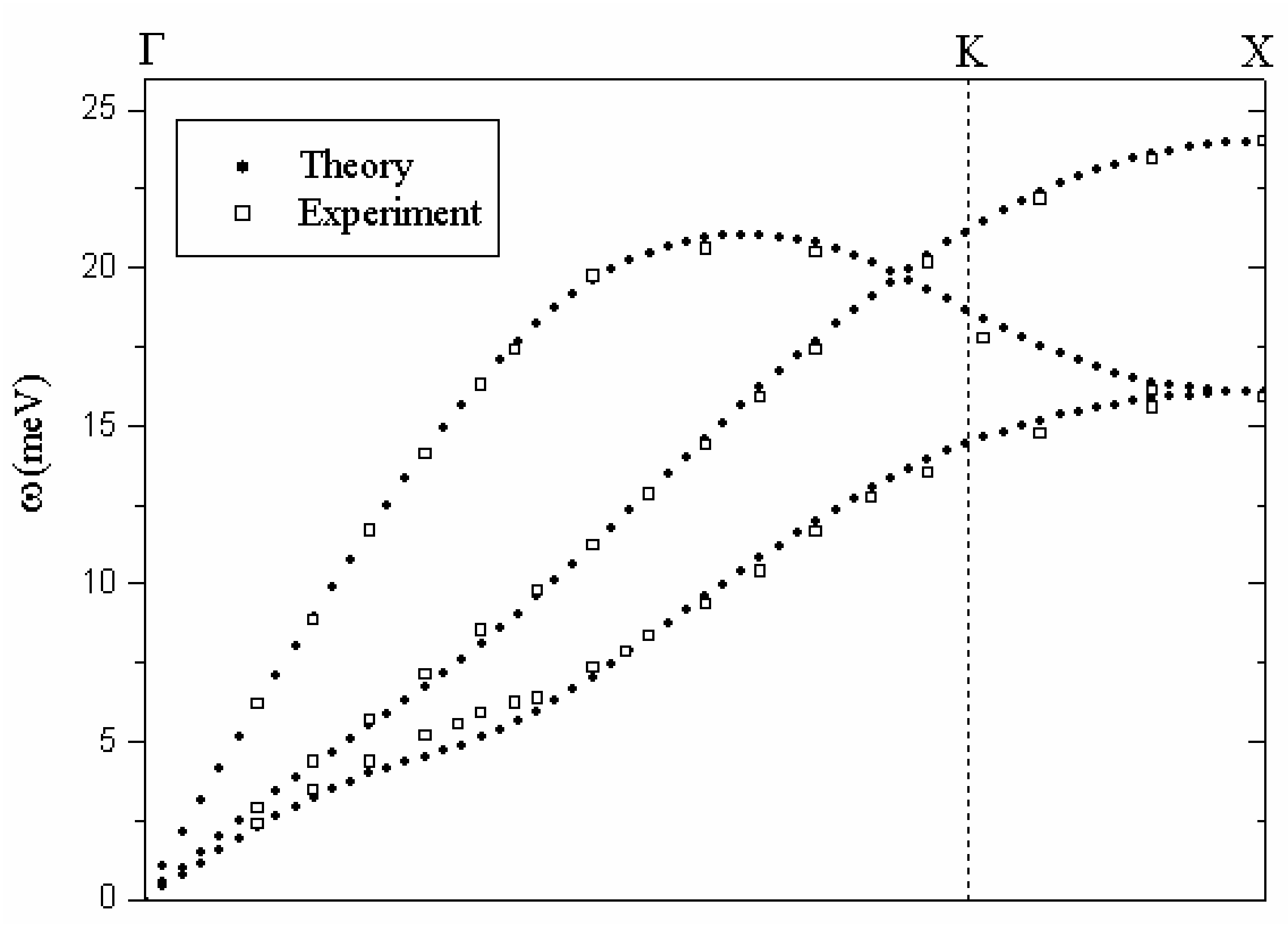

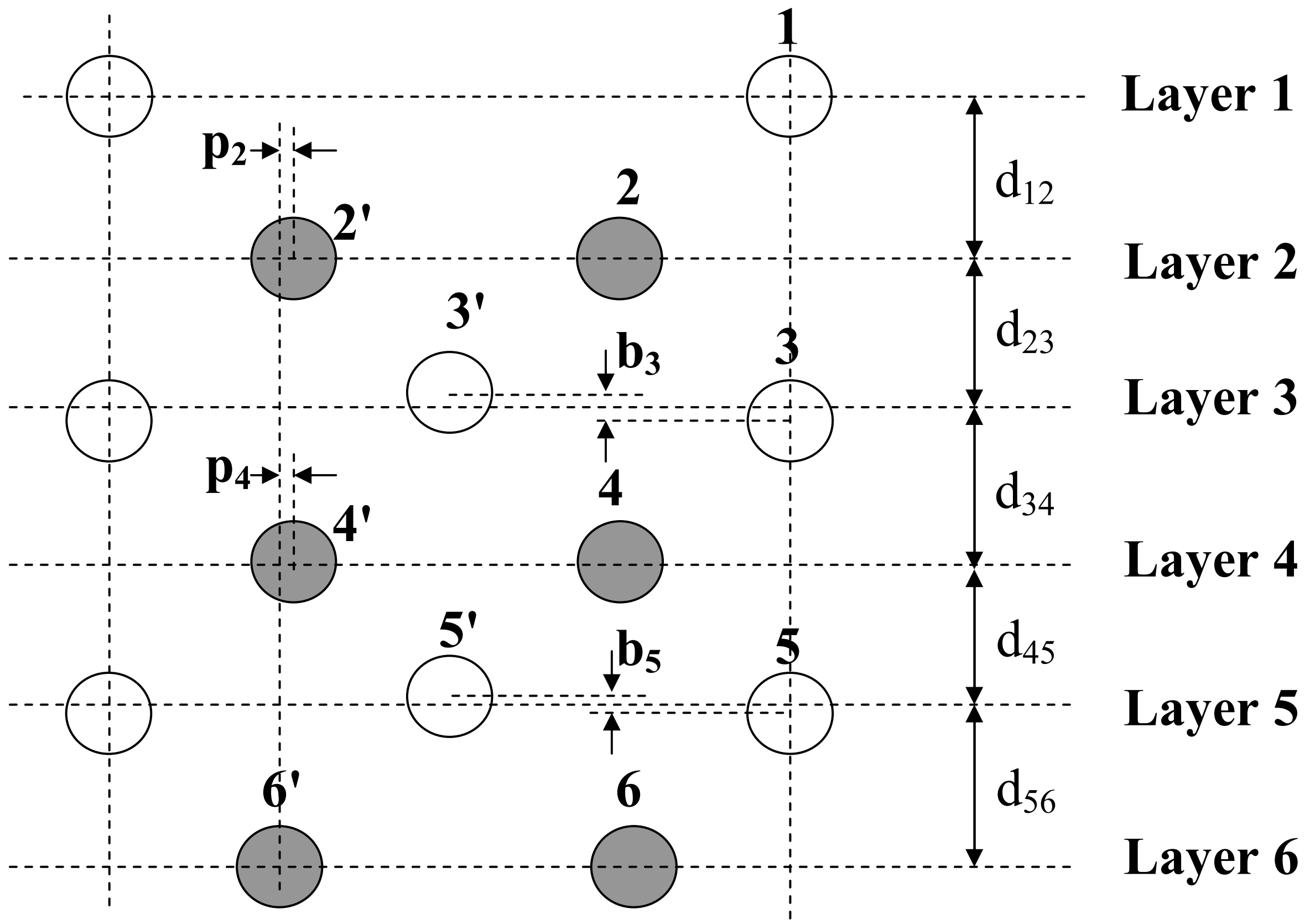

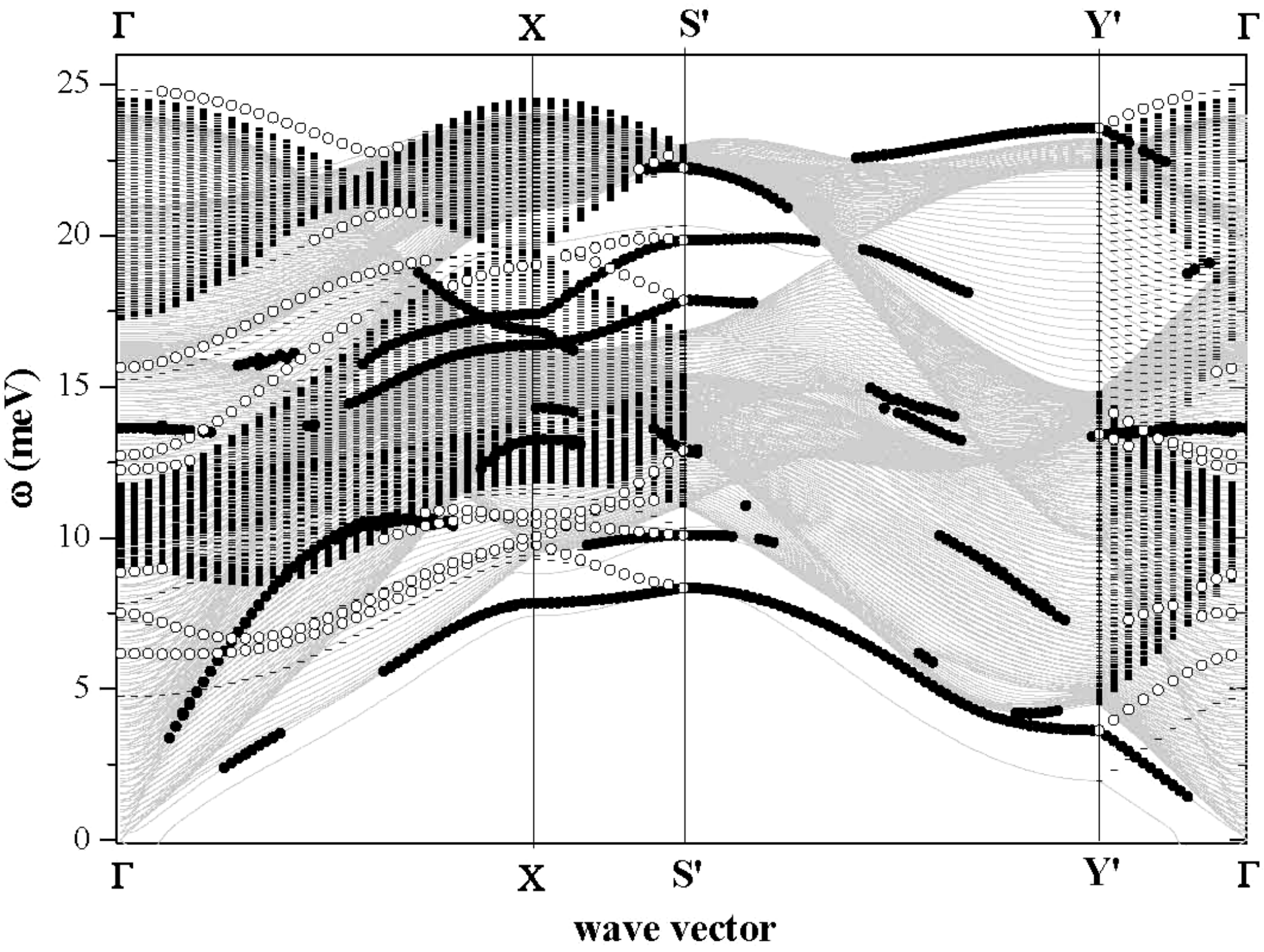

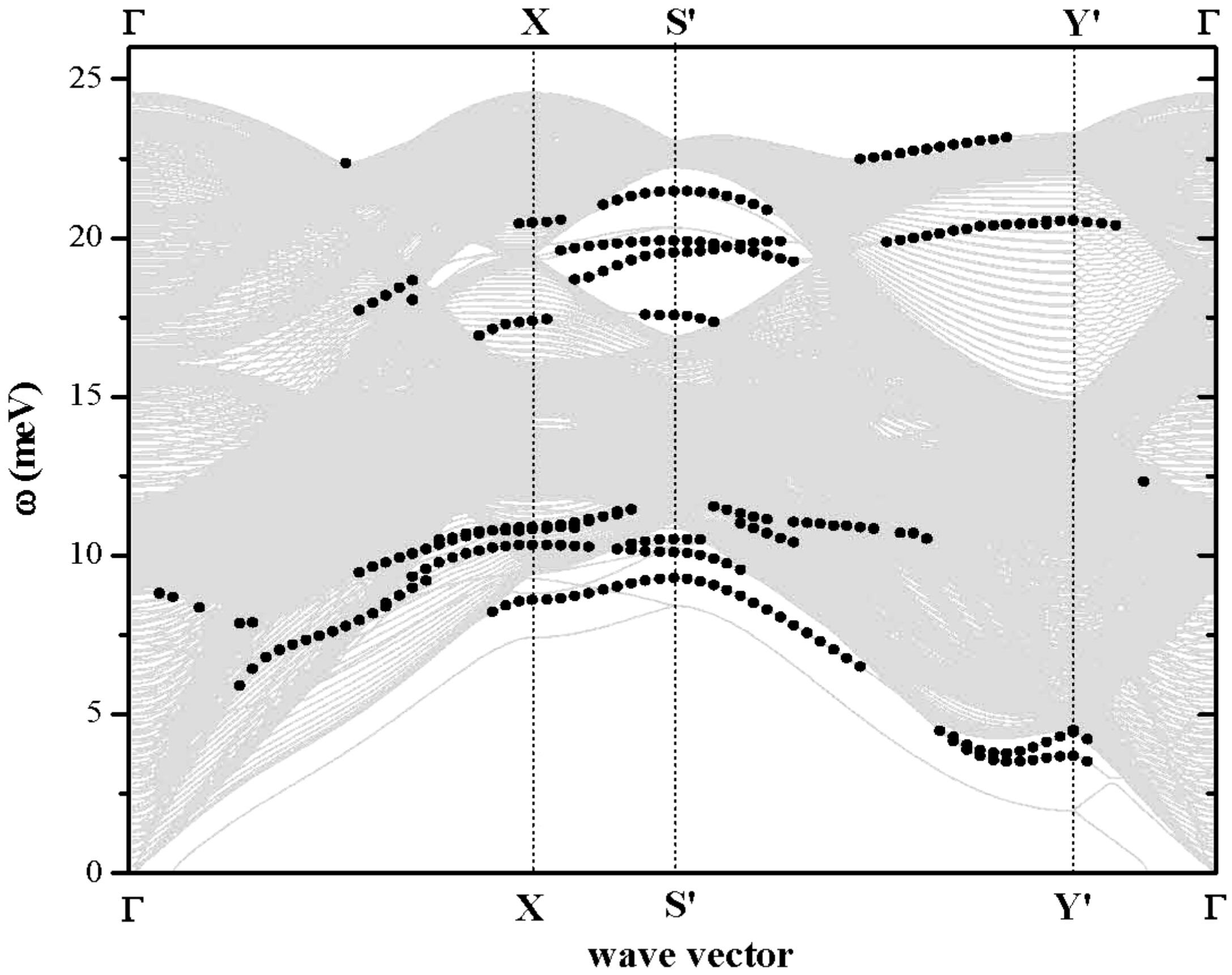

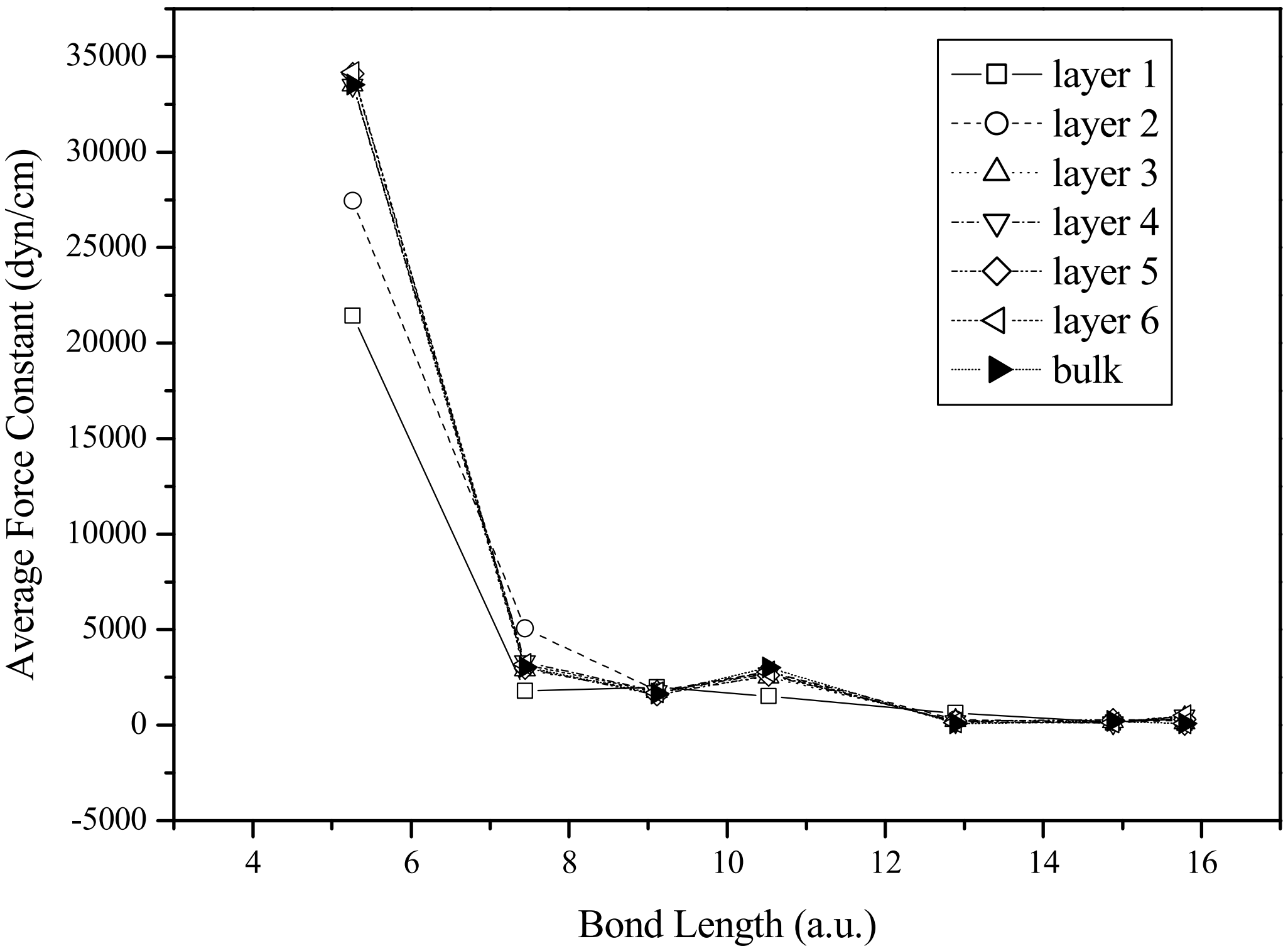

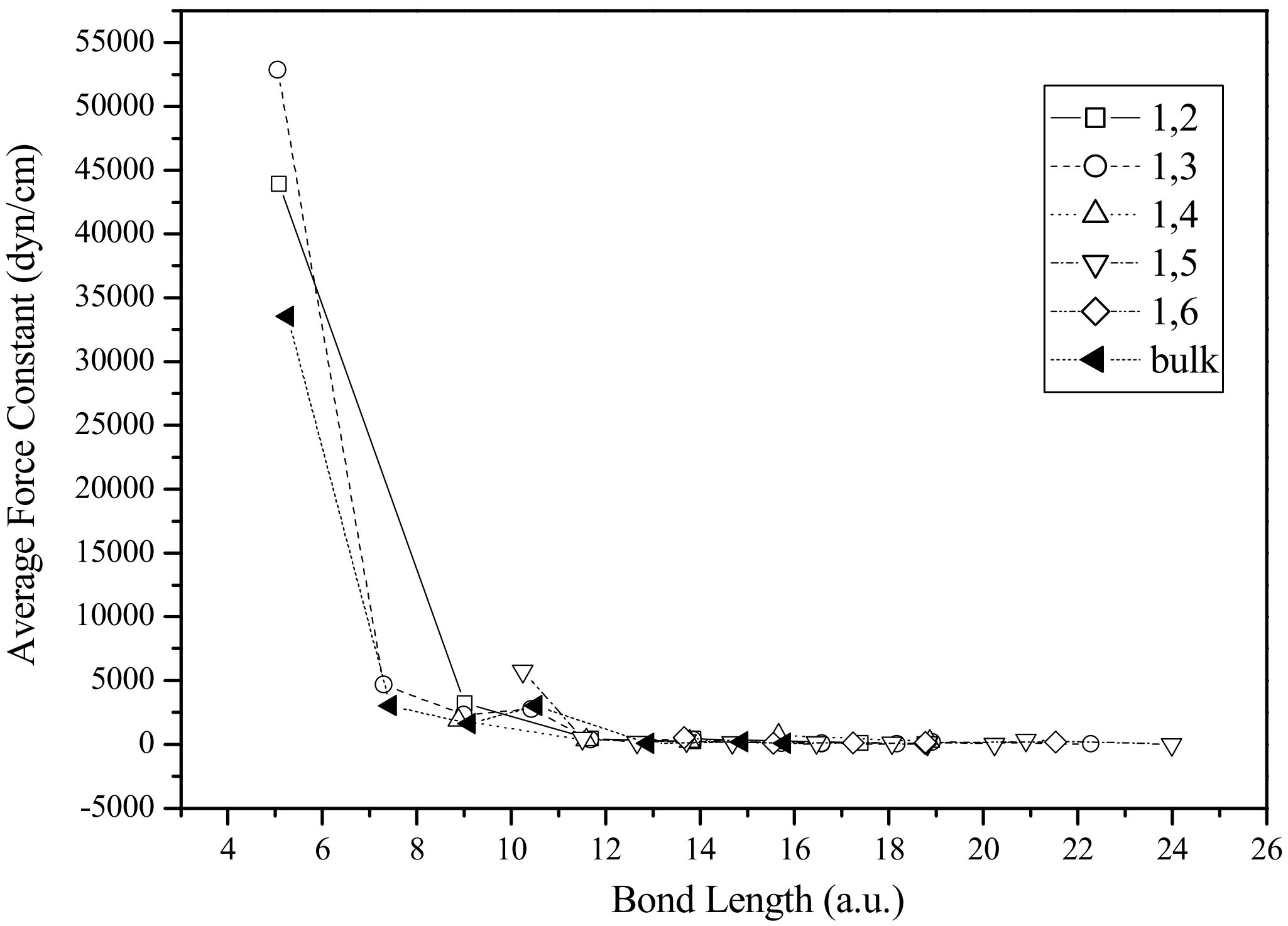

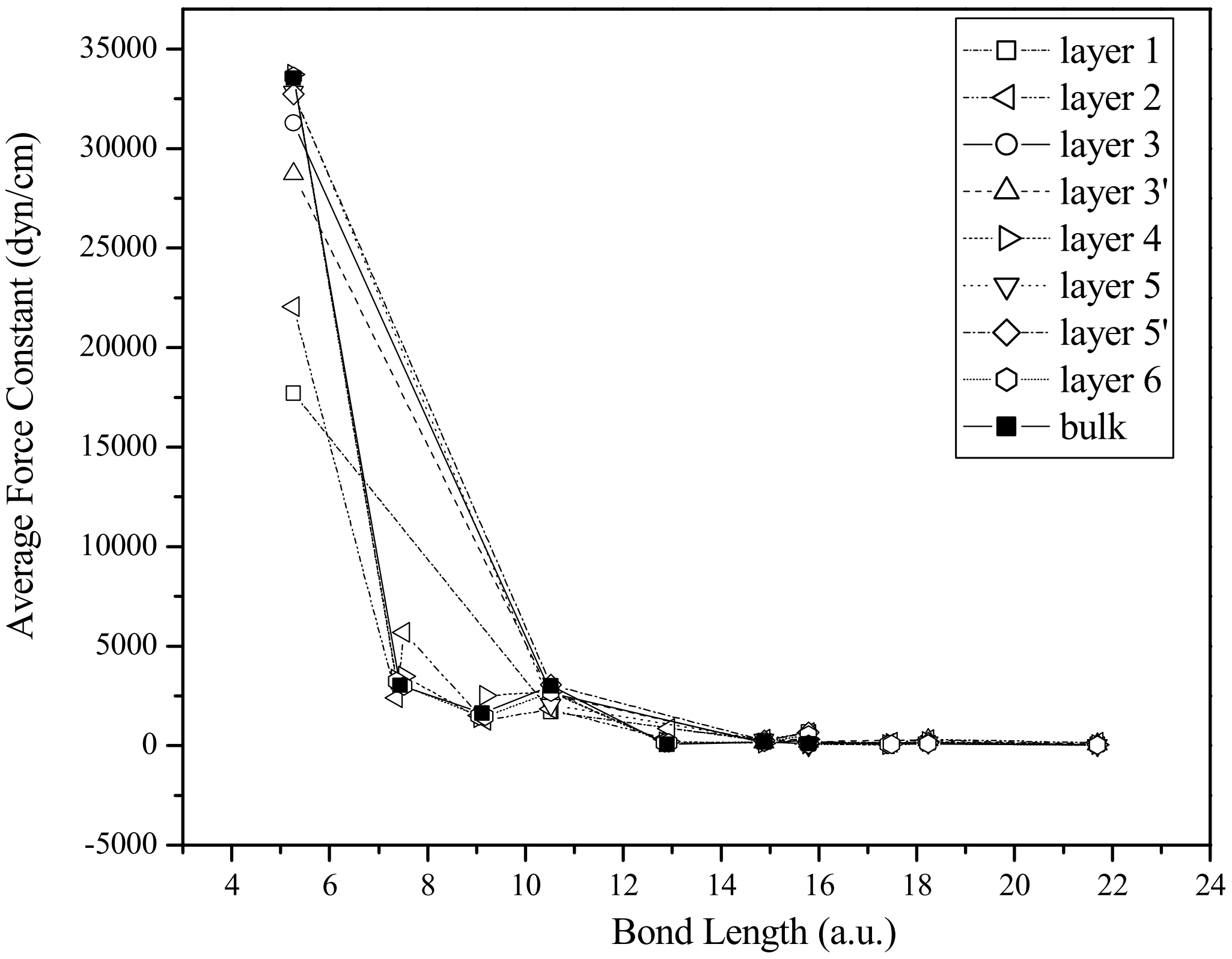

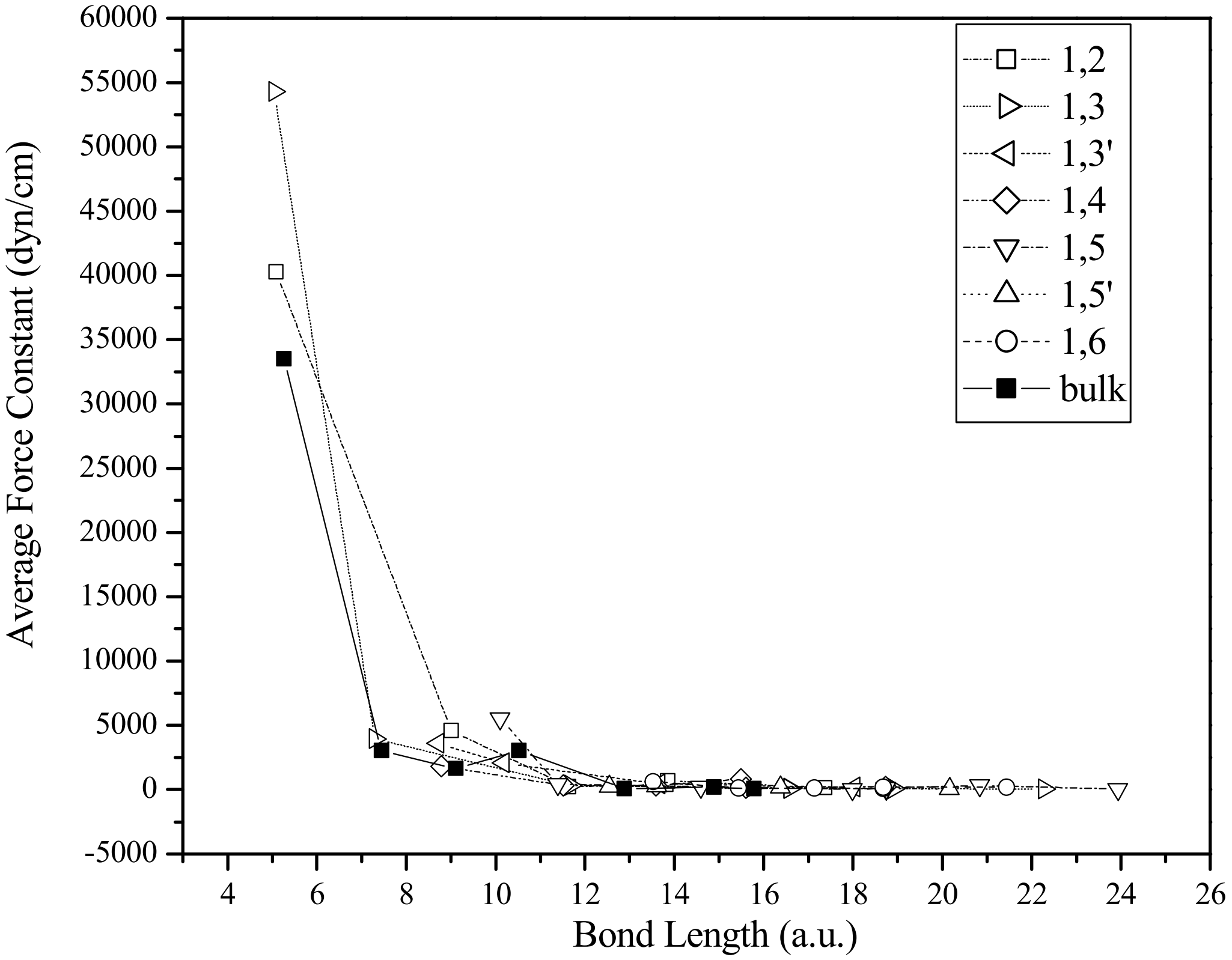

Table I. Calculated structural relaxations for the unreconstructed and reconstructed Pt(110). See Fig. 3.

| Interlayer distance | Unreconstructed Pt(110) | Reconstructed Pt(110) | Experimental results [a] |
|---|---|---|---|
| $d_{12}$ | -14.9% | -16.9% | -16.0 ~ -30.3% |
| $d_{23}$ | +7.0% | -1.0% ($p_2$: 0.02Å) | -0.7 ~ -12.6% ($p_2$:0.04~0.09Å) |
| $d_{34}$ | -2.2% | +1.4% ($b_3$: 0.07 Å) | ($b_3$:0.04~0.32 Å) |
| $d_{45}$ | -0.7% | +0.6% ($p_4$: 0.06 Å) | ($p_4$:0.05~0.12Å) |

[a] Reference 16.

Table II. The force constant with interatomic distance in bulk Pt.

| NN | Force constant(dyne/cm) | | | Bond-length (a.u.) |
|---|---|---|---|---|
| 1 | -27120.6 | 30559.1 | 0.0 | 5.26229 |
| | 30559.1 | -27120.6 | 0.0 | |
| | 0.0 | 0.0 | 5832.6 | |
| 2 | 695.7 | 0.0 | 0.0 | 7.44200 |
| | 0.0 | 695.7 | 0.0 | |
| | 0.0 | 0.0 | -5155.6 | |
| 3 | 60.5 | 917.9 | -917.9 | 9.11455 |
| | 917.9 | 60.5 | 1041.8 | |
| | -917.9 | 1041.8 | -1532.9 | |

Table III. The variation of the average force constant with interatomic distance in bulk Pt.

| NN | Average force constant (dyne/cm) | Bond-length (a.u.) |
|---|---|---|
| 1 | 33530.1 | 5.26229 |
| 2 | 3030.3 | 7.44200 |
| 3 | 1622.6 | 9.11455 |
| 4 | 3016.6 | 10.52458 |
| 5 | 121.4 | 11.76684 |
| 6 | 77.5 | 12.88992 |
| 7 | 204.1 | 13.92271 |
| 8 | 194.2 | 14.88400 |
| 9 | 85.2 | 15.78687 |

Table IV. Intralayer surface force constant matrix.

| Atoms (i, j) | Bond Length (a.u.) | Unreconstructed Pt(110) Force constant (dyne/com) | Reconstructed Pt(110) Force constant (dyne/com) |
|---|---|---|---|
| 1,1 | 5.26229 | -19003.0 25473.2 0.0<br>2310.4 -19003.0 0.0<br>0.0 0.0 -1015.0 | -16995.3 18468.6 0.0<br>4444.5 -16995.3 0.0<br>0.0 0.0 -1097.2 |
| 2,2 | 5.26229 | -21487.9 25984.1 0.0<br>25617.7 -21487.9 0.0<br>0.0 0.0 2142.4 | -18418.2 25217.2 -1590.0<br>10169.8 -18418.2 4009.9<br>4012.3 -1587.7 1496.7 |
| 3,3 | 5.26229 | -27618.3 31100.9 0.0<br>28994.8 -27618.3 0.0<br>0.0 0.0 5897.7 | -25897.8 28569.4 0.0<br>27198.9 -25897.8 0.0<br>0.0 0.0 6316.8 |
| 3',3' | 5.26229 | | -23560.7 28523.0 0.0<br>23421.6 -23560.7 0.0<br>0.0 0.0 1728.1 |
| 4,4 | 5.26229 | -27326.1 30232.4 0.0<br>30445.1 -27326.1 0.0<br>0.0 0.0 5718.9 | -27321.0 29932.5 -933.2<br>31295.3 -27321.0 1390.9<br>1393.9 -930.1 5940.0 |

Table V. Interlayer surface force constant matrix.

| Atoms (i, j) | Unreconstructed Pt(110) | | Reconstructed Pt(110) | |
|---|---|---|---|---|
| | Bond Length (a.u.) | Force constant (dyne/com) | Bond Length (a.u.) | Force constant (dyne/com) |
| 1,2 | 5.07749 | 7136.6 5971.7 -9224.4<br>-284.1 -32044.3 33297.8<br>-3149.7 46905.5 -35573.6 | 5.08174 | 2866.4 5011.9 9697.5<br>-2577.7 -27901.1 -26471.4<br>2720.4 -44705.3 -35305.1 |
| 1,3 | 5.05438 | -41302.9 -49274.1 0.0<br>-49274.1 -41302.9 0.0<br>0.0 0.0 10920.3 | 5.04848 | -43468.9 -49541.2 0.0<br>-49541.2 -43468.9 0.0<br>0.0 0.0 12137.6 |
| 2,3 | 5.35689 | 5871.8 953.5 1296.5<br>200.6 -20663.2 26192.3<br>-688.3 21829.8 -20783.0 | 5.40793 | 5872.2 -1870.6 2118.3<br>-1015.0 -19083.7 25456.1<br>614.3 22972.9 -18562.0 |
| 2,3' | | | 5.09647 | 4769.8 4696.6 7444.8<br>-1547.0 -36547.8 -32082.6<br>2184.5 -41273.1 -34105.2 |
| 2,4 | 5.39017 | -19739.7 -26366.9 0.0<br>-26366.9 -19739.7 0.0<br>0.0 0.0 5479.5 | 5.27155 | -25517.1 -32134.2 -126.0<br>-32134.2 -25517.1 -126.0<br>1752.0 1752.0 7207.8 |
| 3,4 | 5.23424 | 6094.3 673.5 -869.5<br>515.3 -26598.4 31140.4<br>-547.9 32743.1 -28602.9 | 5.23806 | 4273.8 2462.3 2558.2<br>240.4 -25005.2 -30642.2<br>746.0 -31786.4 -29539.0 |
| 3,5 | 5.18705 | -33492.1 -35520.7 0.0<br>-35520.7 -33492.1 0.0<br>0.0 0.0 6404.4 | 5.05589 | -44303.5 -47112.4 0.0<br>-47112.4 -44303.5 0.0<br>0.0 0.0 9238.9 |
| 3',4 | | | 5.33921 | 6416.2 -1393.8 2691.1<br>-2717.2 -22496.4 24378.1<br>1722.9 22701.1 -19601.7 |
| 3',5' | | | 5.57767 | -12686.5 -18449.4 0.0<br>-18449.4 -12686.5 0.0<br>0.0 0.0 2791.3 |
| 4,5 | 5.25297 | 6309.6 112.2 -254.2<br>-115.9 -27702.0 31764.8<br>-113.1 31123.5 -27571.7 | 5.35273 | 4864.5 65.3 -346.8<br>-1023.5 -20324.4 25801.5<br>161.7 23879.3 -21474.4 |
| 4,5' | | | 5.19005 | 6126.4 -208.9 34.6<br>-658.4 -33378.2 -32496.6<br>-75.6 -35033.6 -29915.5 |
| 4,6 | 5.24361 | -28483.3 -33215.1 0.0<br>-33215.1 -28483.3 0.0<br>0.0 0.0 7091.3 | 5.28044 | -26815.6 -29476.0 1501.8<br>-29476.0 -26815.6 1501.8<br>651.8 651.8 5640.1 |